# Stars and the holographic upper bound on gravitational action

Scott Funkhouser
National Oceanic and Atmospheric Administration, 2234 South Hobson Ave.,
Charleston, SC, 29405-2413

ABSTRACT
The holographic upper bound on entropy is applied to the gravitational action associated with the non-relativistic contraction of a nebula. A critical radius is identified, as a function of the initial radius and mass, for which the number of bits associated with the action would equal the maximum number of bits allowed to the body. The gravitational action of a typical star approximately saturates the holographic bound, perhaps suggesting a physical link between holographic principles and astrophysical processes.

Consider an isolated, cold, spherical nebula of gas whose initial radius is $r_0$ and whose mass is $M$. The nebula may have some non-vanishing, uniform rotational angular momentum whose magnitude is $J$. Let the body be subject only to internal forces, and let the mutual gravitational attraction among the constituent particles be, initially, the only significant force. (At some point the contraction may end when another force or effective force becomes significant with respect to the gravitational binding force.) At some initial time $t=0$ let the body begin to contract gravitationally, and let the rotational angular momentum and mass of the body remain constant. There is associated with the contraction a characteristic quantity of action. In this present work the characteristic action of a contracting nebula is examined in scenarios where relativistic effects are not important. (Numerical or geometrical coefficients of order near $10^0$ are also unimportant here.)

Consider times $t$ in which the characteristic radius $r$ of the nebula is much smaller than $r_0$. The magnitude $U$ of the gravitational potential energy of the contracting body will have changed by an amount

$$\Delta U \sim GM\left(\frac{1}{r}-\frac{1}{r_0}\right) \sim \frac{GM}{r}, \tag{1}$$

where $G$ is the Newtonian gravitational coupling. The elapsed time $\Delta t = t$ between the beginning of the contraction and any time $t$ when $r \ll r_0$ is approximately

$$\Delta t \sim \left(\frac{r_0}{g}\right)^{1/2}, \tag{2}$$

where $g \sim GM/r_0^2$ is the initial gravitational acceleration. The action $A$ associated with the contraction is well represented by $\Delta U \Delta t$, and it follows from (1) and (2) that $A$ is given by

$$A \sim \left(\frac{GM^3 r_0^3}{r^2}\right)^{1/2}, \tag{3}$$

for $r \ll r_0$.

Whereas principles of quantum mechanics establish a minimum possible scale of action, the thermodynamics of black holes and holographic considerations lead to a maximum. The holographic upper bound on the number $N(R)$ of degrees of freedom that the contents of a sphere of radius $R$ could occupy is given by

$$N(R) \sim \frac{R^2}{l_P^2}, \qquad (4)$$

where $l_P$ is the Planck length. The maximum entropy of the system and the maximum number of bits of information that could be registered by the system are both proportional to $N(R)$. Since the minimum possible quantum of action is the Planck quantum $\hbar$, there must be associated with any quantity $B$ of action at least $B/\hbar$ registered bits. The maximum action $a(R)$ allowed to the contents of a sphere of radius $R$ is therefore given by

$$a(R) \sim \hbar N(R) \sim \frac{R^2 c^3}{G}. \qquad (5)$$

The holographic limit on action leads to a critical condition for the contracting nebula. As the body contracts and $r$ decreases, the action (3) increases while the upper bound (5) decreases. The gravitational action of the body would be equal to the maximum action allowed to the body if the radius $r$ were to become as small as the critical radius $r_c$ that satisfies the condition

$$\left(\frac{GM^3 r_0^3}{r_c^2}\right)^{1/2} \sim \frac{r_c^2 c^3}{G}, \qquad (6)$$

and thus

$$r_c \sim \left(\frac{GM}{c^2} r_0\right)^{1/2}. \qquad (7)$$

The critical action $A_c$ of the body when $r=r_c$ is given by
$$A_c \sim M r_0 c. \qquad (8)$$
It is instructive to note that the right side of (8) is the holographic upper bound on the rotational angular momentum of a mass $M$ and radius $r_0$ [2]. If rotational angular momentum is conserved during contraction then the $M r_0 c$ represents the maximum possible rotational angular momentum of the body in its final state as well.

As a nebula of mass $M$ and initial radius $r_0$ contracts there must be some mechanism that prevents the body from contracting to radii smaller than $r=r_c$. Note that there is no known mechanism by which holographic principles alone may generate some reaction by which a violation of the holographic bound is prevented. The final, characteristic radius of a contracting body such as the nebula in the present scenario is determined, in general, by one of three primary processes: (a) The body becomes virialized, whereby further contraction is prevented by centrifugal tendencies. (b) The density of the body becomes of the order the characteristic atomic density, whereby contraction is halted by inter-atomic repulsion. (c) The thermal radiation pressure within the body becomes as great as the gravitational pressure. If violation of the holographic upper bound is to be prevented by such mechanisms then there must be some implicit physical connections among astrophysical processes, the physical parameters of astronomical structures and holographic principles.

Consider a star whose mass is of the order $10^{30}$kg that forms from the gravitational contraction of a nebula whose initial characteristic radius is of the order $10^{13}$m, or hundreds of astronomical units. The critical radius at which the action associated with the gravitational contraction will saturate the holographic upper bound is of the order $10^8$m, which is of the order the characteristic radius of a solar-mass star.

Stars may therefore constitute a class of bodies for which the holographic upper bound is nearly saturated by the gravitational action. Typical stars are also remarkable since they represent the unique situation in which all three of the processes (a) – (c), listed above, are comparably significant in preventing further gravitational contraction.